\documentclass[aps,pre,twocolumn,amsmath,amssymb,notitlepage,noeprint,superscriptaddress]{revtex4-1}
\pdfoutput=1
\usepackage[latin9]{inputenc}
\usepackage{braket}
\usepackage{xcolor}
\usepackage{verbatim}
\usepackage{graphicx}
\usepackage{esint}
\usepackage{float}
\definecolor{darkblue}{rgb}{0,0,0.6}
\usepackage[colorlinks,linkcolor=darkblue,citecolor=darkblue,urlcolor=darkblue]{hyperref}

\def\ur{{\underline r}}
\def\br{\mathbf{r}}
\def\upi{{\underline \pi}}
\def\uv{{\underline v}}

\begin{document}

\title{Relationship between two-level systems and quasi-localized normal modes in glasses}

\author{Dmytro Khomenko}

\affiliation{Department of Chemistry, Columbia University, New York, NY 10027, USA}

\author{David R. Reichman}

\affiliation{Department of Chemistry, Columbia University, New York, NY 10027, USA}

\author{Francesco Zamponi}  

\affiliation{Laboratoire de Physique de l'Ecole Normale Sup\'erieure, ENS, Universit\'e PSL, CNRS, Sorbonne Universit\'e, Universit\'e de Paris, 75005 Paris, France
}


\begin{abstract}
Tunnelling Two-Level Systems (TLS) dominate the physics of glasses at low temperatures. Yet TLS are extremely rare and it is extremely difficult to directly observe them {\it in silico}. It is thus crucial to develop simple structural predictors that can provide markers for determining if a TLS is present in a given glass region.
It has been speculated that Quasi-Localized vibrational Modes (QLM) are closely related to TLS, and that one can extract information about TLS from QLM. In this work we address this possibility. In particular, we investigate the degree to which a linear or non-linear vibrational mode analysis can predict the location of TLS independently found by energy landscape exploration.  We find that even though there is a notable spatial correlation between QLM and TLS, in general TLS are strongly non-linear and their global properties cannot be predicted by a simple normal mode analysis. 
\end{abstract}

\maketitle

\section{Introduction}
 
The thermodynamic behavior of low-temperature amorphous solids has been a topic of considerable interest since the seminal experiments of Zeller and Pohl demonstrated marked deviations between the thermal properties of glasses and those of crystalline solids, nearly fifty years ago~\cite{Zeller_and_Pohl_prb_1971}.  In particular, in the range of one degree Kelvin, these experiments and numerous experiments that have followed~\cite{loponen1982,BM88,boiron1999,burin2013,queen2013excess,perez2014two,perezcastaneda2014,liu2014,Queen2015} have demonstrated that the specific heat of a disordered solid is much larger than that of a crystal composed of the same material, with a temperature dependence that varies linearly, as opposed to cubically, and a thermal conductivity that varies quadratically, as opposed to cubically, with temperature.  A theoretical framework provided independently by Anderson, Halperin and Varma~\cite{anderson1972anomalous} and by Phillips~\cite{phillips1972tunneling,phillips87}, ascribes the origin of this puzzling behavior to dilute defects which tunnel between their lowest lying quantum states at low temperatures.  This two-level systems (TLS) picture has successfully rationalized diverse experimentally observed properties, although several outstanding puzzles lie beyond its reach~\cite{phillips87,galperin1989,leggett2013}.  In particular, understanding the microscopic nature of the TLS, as well as the quasi-universal aspects of the thermodynamic data, have remained as outstanding challenges~\cite{zhou2019universal,CY20,ABS20}.

Over the last five years, the swap Monte Carlo technique has provided a major advance in computational glass physics, opening the door to the creation of {\em in silico} glasses 
that have comparable stability properties to those found in the laboratory~\cite{ninarello2017models}.  
Using this technique, and building on previously developed landscape exploration algorithms~\cite{SW82,SW85,heuer1993microscopic,dab1995low,heuer1996collective,DVR99,reinisch2004local,reinisch2005moving,He08,damart2018atomistic}, we recently provided a detailed computational investigation of the TLS model~\cite{Khomenko2020}, providing a direct microscopic description of TLS and 
demonstrating that their density decreases as the cooling rate decreases, similar to what is seen in several recent experiments~\cite{queen2013excess,perez2014two,perezcastaneda2014,liu2014,Queen2015}.  Although this work only considered one model system, and thus the question of the quasi-universality ~\cite{BM88} of low-temperature thermodynamic anomalies could not be investigated, a detailed description of the nature of TLS was provided.  Specifically, tunneling motion in TLS was found to be comprised of defect-vacancy-like motion of one or a handful of particles, although occasionally highly collective tunneling motion of a large number of particles was observed.

A major issue with the landscape exploration methods currently used to identify TLS {\it in silico}~\cite{heuer1993microscopic,DVR99,damart2018atomistic,Khomenko2020}
is that they are computationally very expensive. This bottleneck is due to (i)~the need to accumulate a sufficient number of inherent structures (IS)~\cite{SW82,SW85} and then, for at least the most promising pairs of IS (see~\cite{Khomenko2020} for details),
(ii)~the need to identify a relaxation pathway between the two minima in the $3N$-dimensional energy landscape achieved via a computationally expensive minimization of a path function
in the space of possible paths~\cite{jonsson1998nudged,DVR99,henkelman2000climbing,bolhuis2002transition,vanden2010transition}.
Hence, it would be extremely helpful to identify {\it a priori} the glassy configurations (or sub-regions of them) that are most likely to include TLS with the proper energy splitting,
via  some sort of simple structural indicator.

Over the last two decades a seemingly distinct type of (partially) localized entity, namely 
quasi-localized vibrational modes (QLM)~\cite{Schober_Laird_numerics_PRL}, 
have been intensely scrutinized.  QLM are characterized by a defect-like localized core with a power-law decaying elastic background, and are prominently found in the low-frequency wing of the density of states of amorphous systems.  They have been connected to the universal non-Debye behavior of the low-frequency density of states~\cite{BaityJesi2015,Edan2016,modes_prl_2016,MSI17,wang2019low,pinching_pnas,modes_prl_2020,itamar_GPS_prb_2020} and to the attenuation of sound waves in glassy systems~\cite{MMB16,gelin2016anomalous,PAPR19,scattering_jcp}, to the dynamical heterogeneity upon approach to the glass transition from the high temperature side~\cite{Schober_correlate_modes_dynamics,widmer2008irreversible}, to the plasticity of the glass under strain~\cite{manning2011,wyart2019,david_huge_collaboration}, and to the critical-like behavior in jamming~\cite{Masanari2018,atsushi_large_d_packings_pre_2020}.  

It is natural to assume a connection between QLM and TLS. Indeed, it is known from computer simulations that as model supercooled liquids are cooled, the concentration of real-space localized cores associated with QLM rapidly decreases~\cite{wang2019low,pinching_pnas}, as is also the case for TLS.  It is thus reasonable to assume that at the glass transition these QLM cores are ``frozen'' into the sample and provide the seeds for low-temperature defects~\cite{lubchenko2001}.  Indeed, this notion is central to the successful soft potential model~\cite{soft_potential_model_1991,Schober_prb_1992,buchenau1993,GPS03,DHLP20} of low-temperature glasses, which extends the models of Anderson-Halperin-Varma and of Phillips to somewhat higher temperatures by connecting TLS to anharmonic vibrational modes in the glass. If this connection is precise, QLM could be used as structural predictors for the location of TLS in glass samples, thus aiding the computational search for tunneling states. Moreover, establishing this connection more precisely could help validate or invalidate models of low-temperature glasses based on interacting anharmonic modes~\cite{kuhn_and_Horstmann_prl_1997,GPS03,DHLP20,rainone2020solvable,albert2020searching}.

In this work, leveraging our ability to prepare realistically cooled samples and extract detailed information about both TLS and QLM, we explore their putative connection in detail.  In our system we find that, in real space, TLS and QLM are well correlated, in the sense that particles that move the most in a TLS are typically close to particles that move the most in a QLM. On the other hand, in phase space, we demonstrate a surprisingly weak correlation between TLS and both linear and non-linear QLM: 
the $3N$-dimensional vector $\ur_{AB}$ that encodes the displacement in phase space of all particles in a TLS is often completely unrelated to the 
vectors that define QLMs.

 \begin{figure}[t]
  \includegraphics[width=\columnwidth]{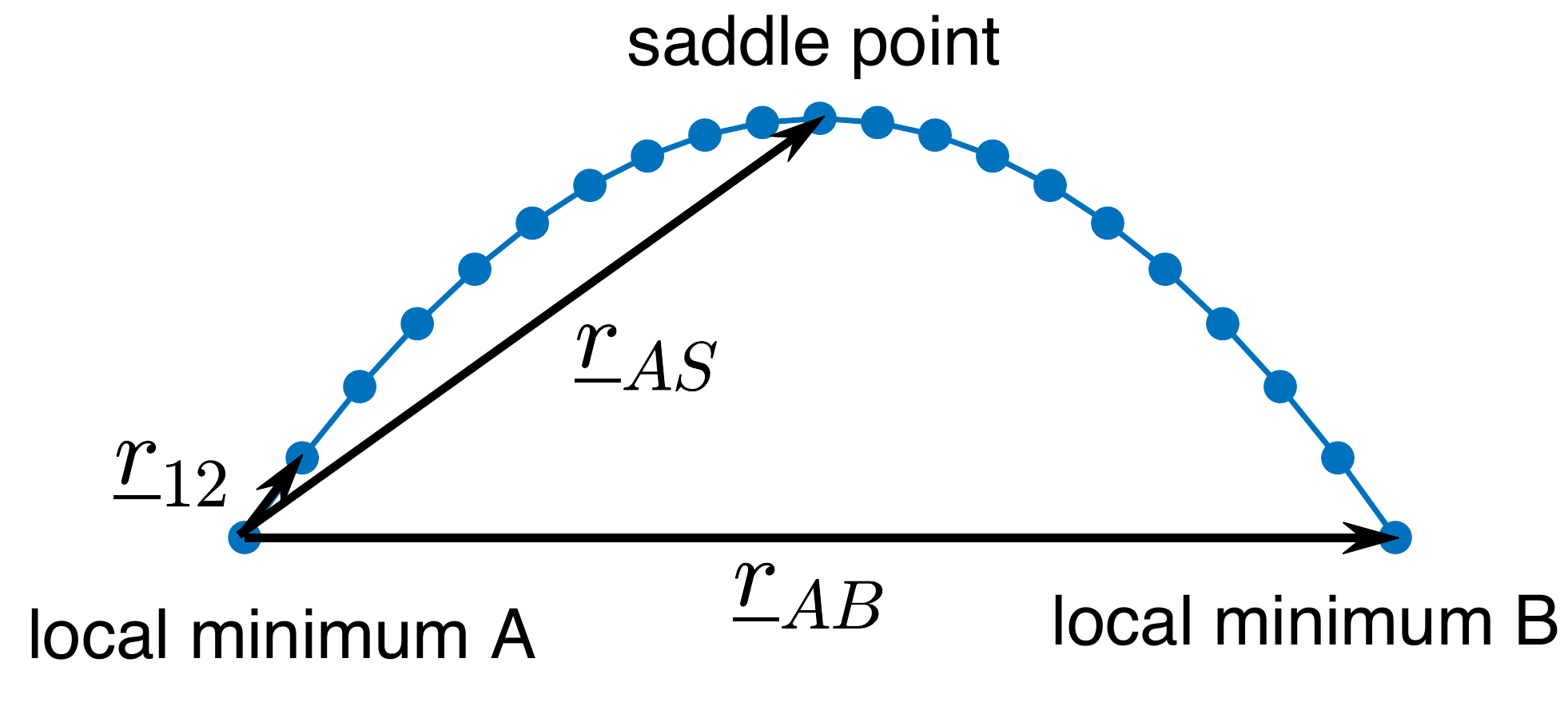}
  \caption{
  Schematic illustration of the multidimensional minimal energy transition path between two energy minima within a double-well potential.
  Also illustrated are the displacement vector $\ur_{AB}$ between the two minima, the displacement vector $\ur_{AS}$ between the first minimum and the saddle point,
  and the tangent vector $\ur_{12}$ in the first minimum.
  } 
  \label{fig:ilu} 
\end{figure}

\section{System} 

Our study is based on our previous work~\cite{Khomenko2020}, in which we prepared {\em in silico} glasses at cooling rates,
from poorly annealed to ultrastable, and explored their energy landscape. 
For completeness, we will give a very brief summary of the methodology, but we refer to Ref.~\cite{Khomenko2020} for details.

Our system is a non-additive polydisperse mixture of $N=1500$ particles.
In the following, $\ur$ denotes a $3N$-dimensional phase space vector encoding the position of all particles, 
$\br_i$ denotes the three-dimensional coordinate of particle $i=1,\cdots,N$, and $r_{ij} = |\br_i - \br_j|$ is the scalar distance
between particles $i$ and $j$. We use the following inter-particle interaction potential:
\begin{equation}\label{eq:potential}
v_{ij}(r_{ij}) =\begin{cases}
 \epsilon \left(\frac{\sigma_{ij}}{r_{ij}}\right)^{12}+ \epsilon F\left(\frac{\sigma_{ij}}{r_{ij}}\right) \ , & r_{ij} < r_{cut}, \\
 0 \ , & r_{ij} > r_{cut}, 
\end{cases}\end{equation}
 where $r_{cut} = 1.25 \sigma_{ij}$, and $\sigma_{ij}$ is the non-additive interaction length scale associated to particle pair $ij$. 
The function $F(x)$ is a fourth-order polynomial which guarantees the continuity of the potential up to its second derivative at~$r_{cut}$. 
We express all dimensional quantities in units of energy $\epsilon$, length $\langle\sigma\rangle=1$ (the average being over particle pairs), time $\sqrt{\epsilon/(m \langle\sigma\rangle^2)}$, and the number density is set to $\rho=1$ in these units. The mode-coupling temperature, which sets
the onset of strongly glassy dynamics, is $T_{\rm MCT} = 0.104$~\cite{ninarello2017models}.

 Using the swap Monte Carlo algorithm~\cite{ninarello2017models} we prepare fully equilibrated configurations at three different preparation temperatures $T_f=0.062$ (ultra-stable glasses), 0.07 (liquid cooled glasses), 0.092 (poorly annealed glasses). 
 Normal molecular dynamics (MD) initialized in these configurations is fully arrested (except for $T_f=0.092$, at which slow residual diffusion is observed), and each configuration thus defines a glass basin in the potential energy landscape.
 The temperature $T_f$ corresponds to Tool's ``fictive temperature''~\cite{tool}, and encodes the degree of glass stability.
  For each of the glasses, we explore the glass basin in the energy landscape and determine a set of local energy minima, or ``inherent structures'' (IS),
by running MD at a slightly lower temperature $T_{\rm MD}=0.04$ and periodically minimizing the system's potential energy~\cite{SW82,SW85}. 
Pairs of energy minima that are subsequently visited a large enough number of times (see~\cite{Khomenko2020} for details)  are further analysed by the 
Nudged Elastic Band (NEB) method~\cite{jonsson1998nudged,henkelman2000climbing} to find the minimum energy pathway connecting them, 
see Fig.~\ref{fig:ilu},
and the associated value of the classical energy barrier. This procedure allows us to obtain a library of distinct Double Well potentials (DWs) in the high-dimensional energy landscape.
For each DW, we then solve an effective one-dimensional Schr\"odinger equation to obtain the quantum tunnel splitting $E = {\cal E}_2 - {\cal E}_1$ 
between the first two energy levels. 
We find that the relevant DWs, which define active TLS in the quantum regime (namely those with a tunnel splitting equal to or below the temperature $T_{Q}$ which defines the low temperature regime in~\cite{Khomenko2020}),
have $E < 0.0015$ for Argon parameters~\cite{DVR99}, and $E<0.0002$ for NiP metallic glass parameters~\cite{heuer1993microscopic},
see~\cite{Khomenko2020} for details.

  \begin{figure}[t]
   \centering{
  \includegraphics[width=\columnwidth]{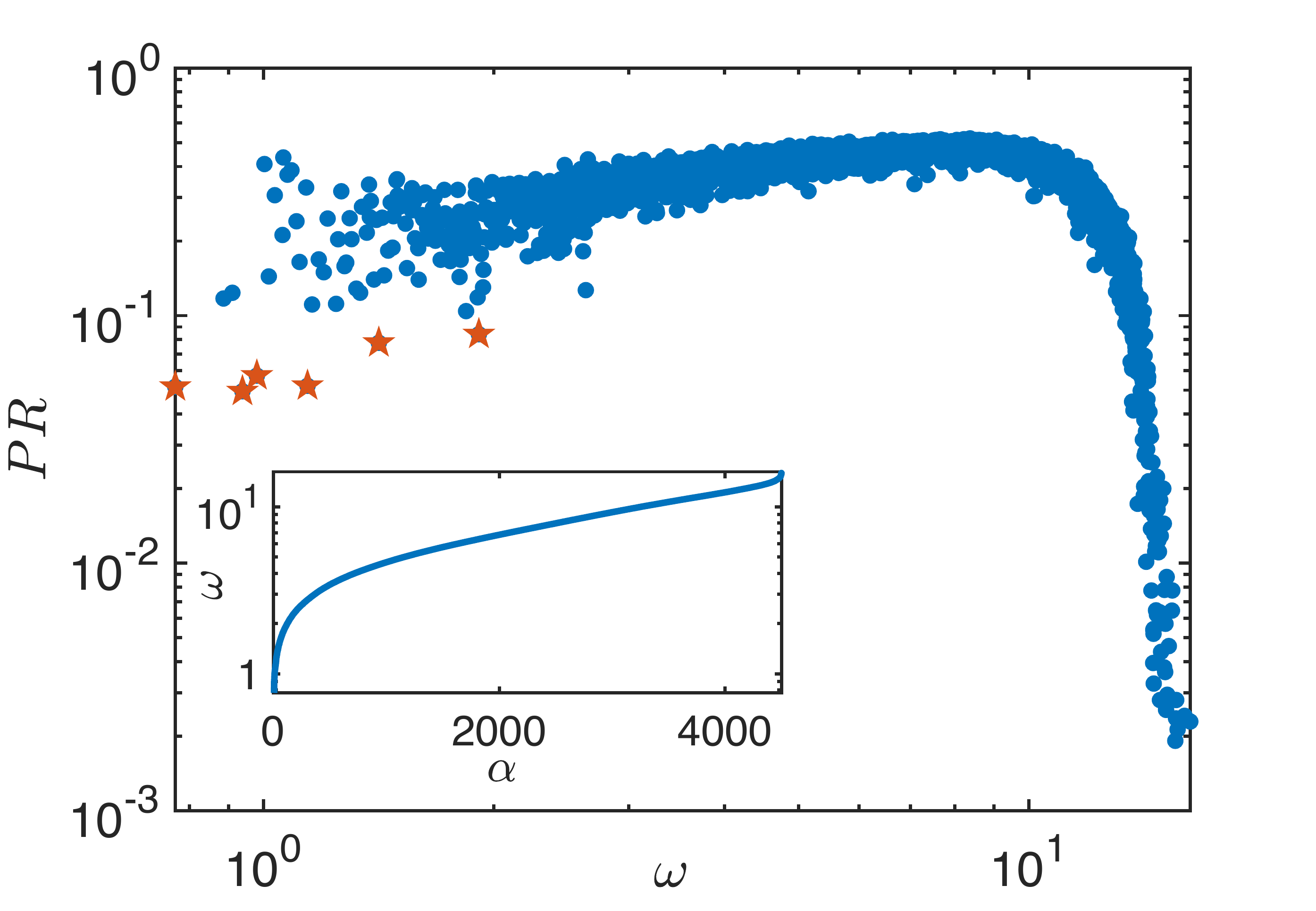}}
  \caption{Participation ratio of normal modes versus their frequency, for one selected energy minimum. Modes shown in red stars are considered to be quasi-localised.
  In the inset, the mode frequency is shown as a function of the mode index.
  } 
  \label{fig:local} 
\end{figure}

\section{Normal modes} 

Having curated a library of DWs along with associated displacement fields $\ur_{AB}$, we can perform a normal mode analysis in each of the two minima $A$ and $B$ (because our search
procedure is statistically symmetric, in the following
we focus on minimum $A$ without loss of generality),
and check if the normal modes overlap with the DW displacement field, to be defined below. We now define more precisely the linear and non-linear normal modes and their relationship with
the minimum energy path illustrated in Fig.~\ref{fig:ilu}.
 
 Several displacement fields can be associated with a DW transition: the difference between the two energy minima, $\ur_{AB}$, the 
 difference between the first minimum and the saddle point, $\ur_{AS}$, and the tangent direction  $\ur_{12}$ of the minimal energy path in $A$ (see Fig.~\ref{fig:ilu}). 
 The vector $\ur_{12}$ is estimated by a discretization of the minimum energy path, as the difference between the position of the first two beads (or images) of the NEB path.

Next, we define the tensors $M$ and $U$ as follows:
 \begin{equation}
M_{\alpha\beta}=\frac{\partial^2 V}{\partial r_\alpha \partial r_\beta} \ , \quad
U_{\alpha\beta\gamma}=\frac{\partial^3 V}{\partial r_\alpha \partial r_\beta \partial r_\gamma} \ ,
\end{equation}
where $V(\ur) = \sum_{i<j} v_{ij}(r_{ij})$ is the total potential energy, Greek indices such as $\alpha=1,\cdots,3N$ run over all the $3N$ coordinates of phase space
vectors, and the particle coordinates are evaluated in minimum $A$ after derivatives are taken.
 The linear vibrational modes $\uv^\alpha$, for $\alpha=1,\cdots,3N$, are defined as the eigenvectors of the Hessian matrix $M$,
 \begin{equation}
 M \uv  = \omega^2 \uv \ ,
 \end{equation}
 and $\omega$ is the corresponding frequency. For a system of size $N=1500$, we can find all the eigenvectors by using standard matrix diagonalization packages.
It has been shown that in glasses, collective excitations such as phonons coexist with QLM, which feature a localized core decorated by a power-law
decaying elastic tail~\cite{Schober_Laird_numerics_PRL,Edan2016,modes_prl_2016,MSI17,scattering_jcp,wang2019low,pinching_pnas,modes_prl_2020}.
Unfortunately,
there is no clear way of separating delocalized phonon modes from QLMs. 
Here, we use an arbitrary threshold on the mode's participation ratio,
 \begin{equation}\label{PR}
     PR(\uv)=\left[N\sum_{i=1}^N |\mathbf{v}_i|^4\right]^{-1},
 \end{equation}
 where $\mathbf{v}_i$ is the displacement of particle $i$ in the mode $\uv$, normalized to $|\uv|=1$. 
In Fig.~\ref{fig:local} we show the participation ratio as a function of the modes' frequency.
We consider low-frequency modes with $PR<0.1$ to be QLMs. 

\begin{figure}[t]
   \begin{tabular}{cc}
      linear mode &non-linear mode\\
      \includegraphics[width=0.5\columnwidth]{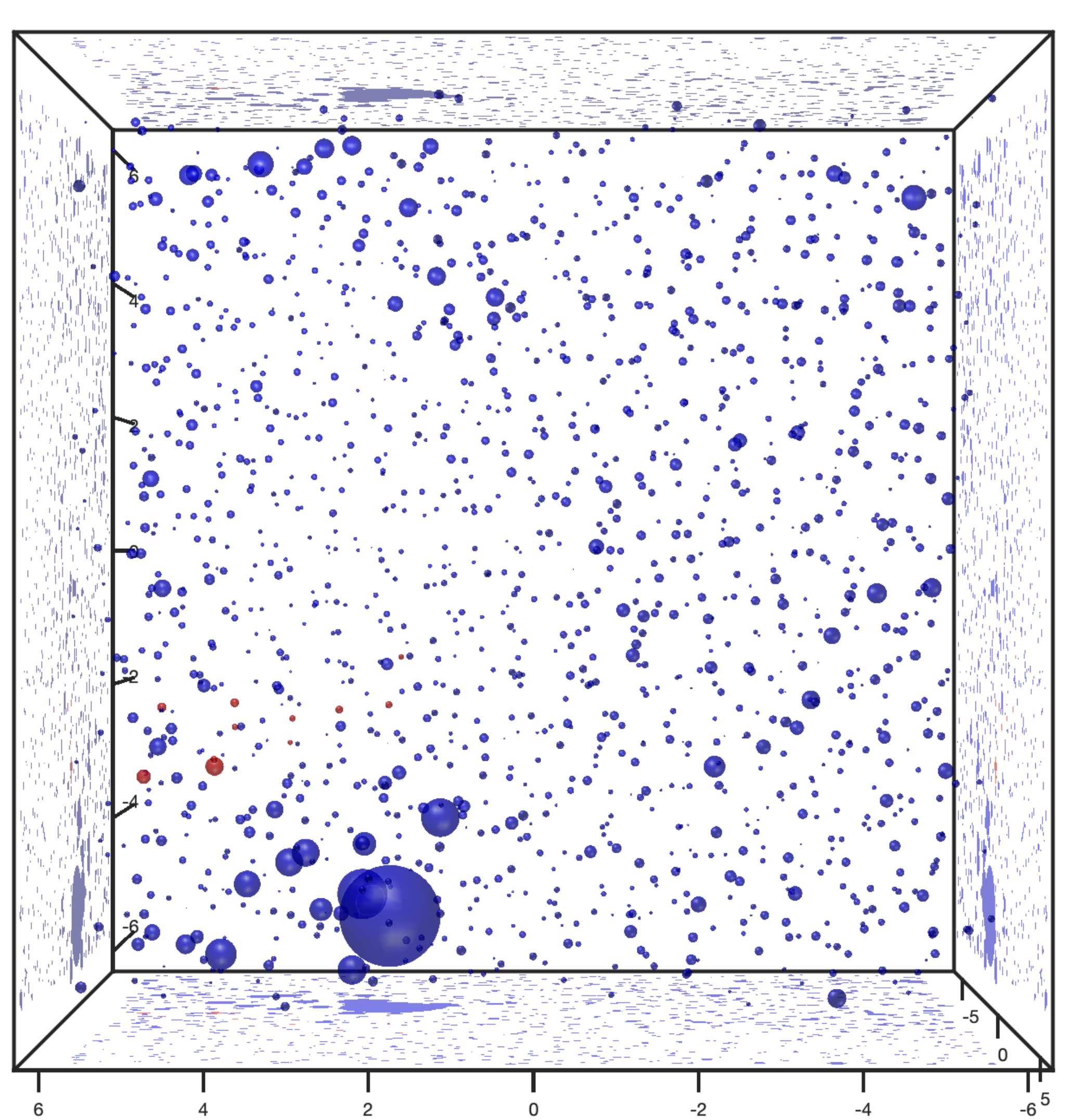}&
      \includegraphics[width=0.5\columnwidth]{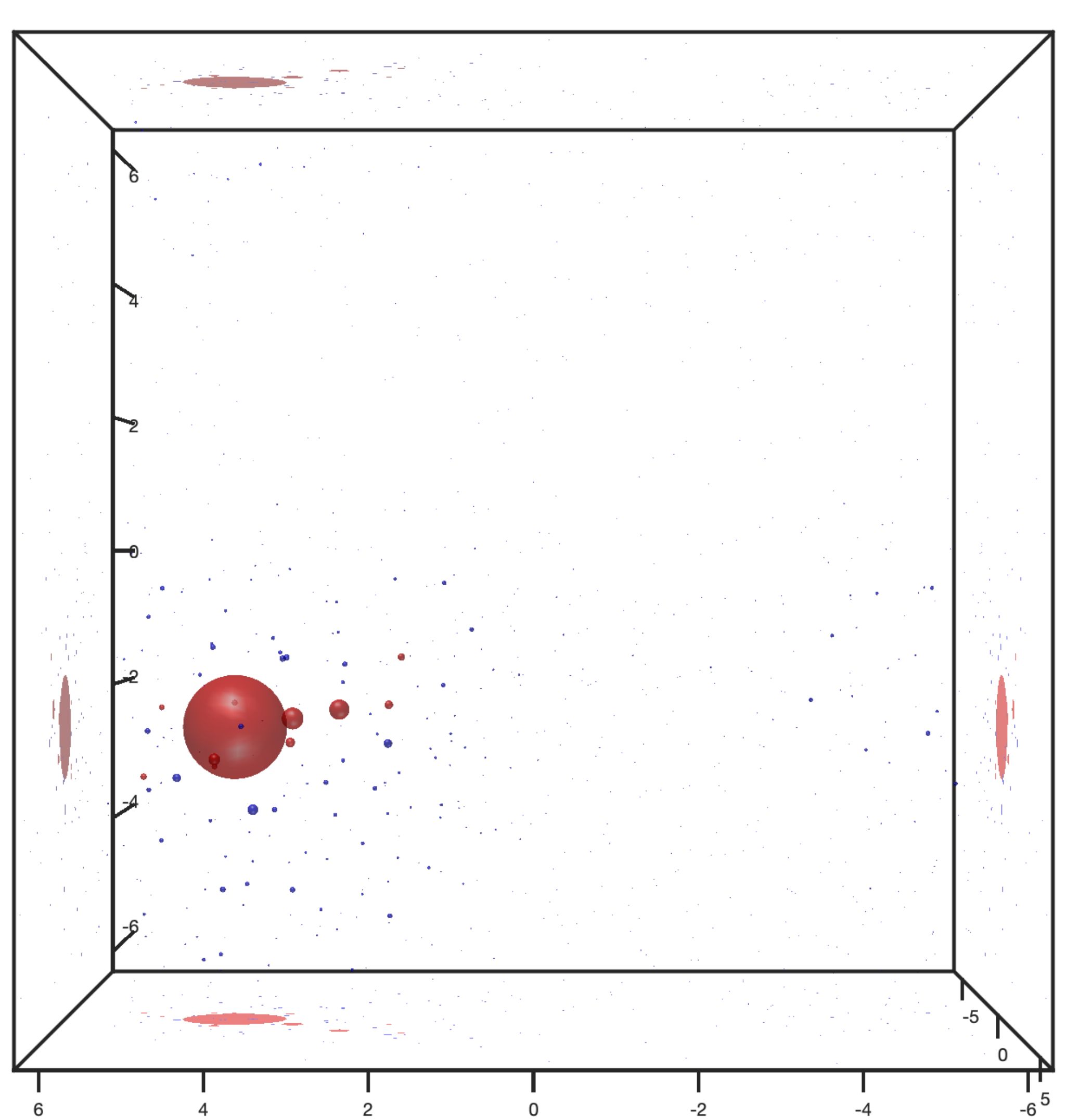}
    \end{tabular}
  \caption{Displacement fields of the linear and the non-linear modes that have the largest overlap with the displacement field $\ur_{AB}$ of a double well potential. 
  The size of the particles is proportional to the particle displacement. The ten particles that move the most in the $\ur_{AB}$ field are shown in red. 
  In the box faces, the corresponding projections are shown.}
  \label{fig:configuration} 
\end{figure}

  \begin{figure*}[t]
\begin{tabular}{ccc}
$N_b=40$ & $N_b=200$ & $N_b=600$\\
\includegraphics[scale=0.3]{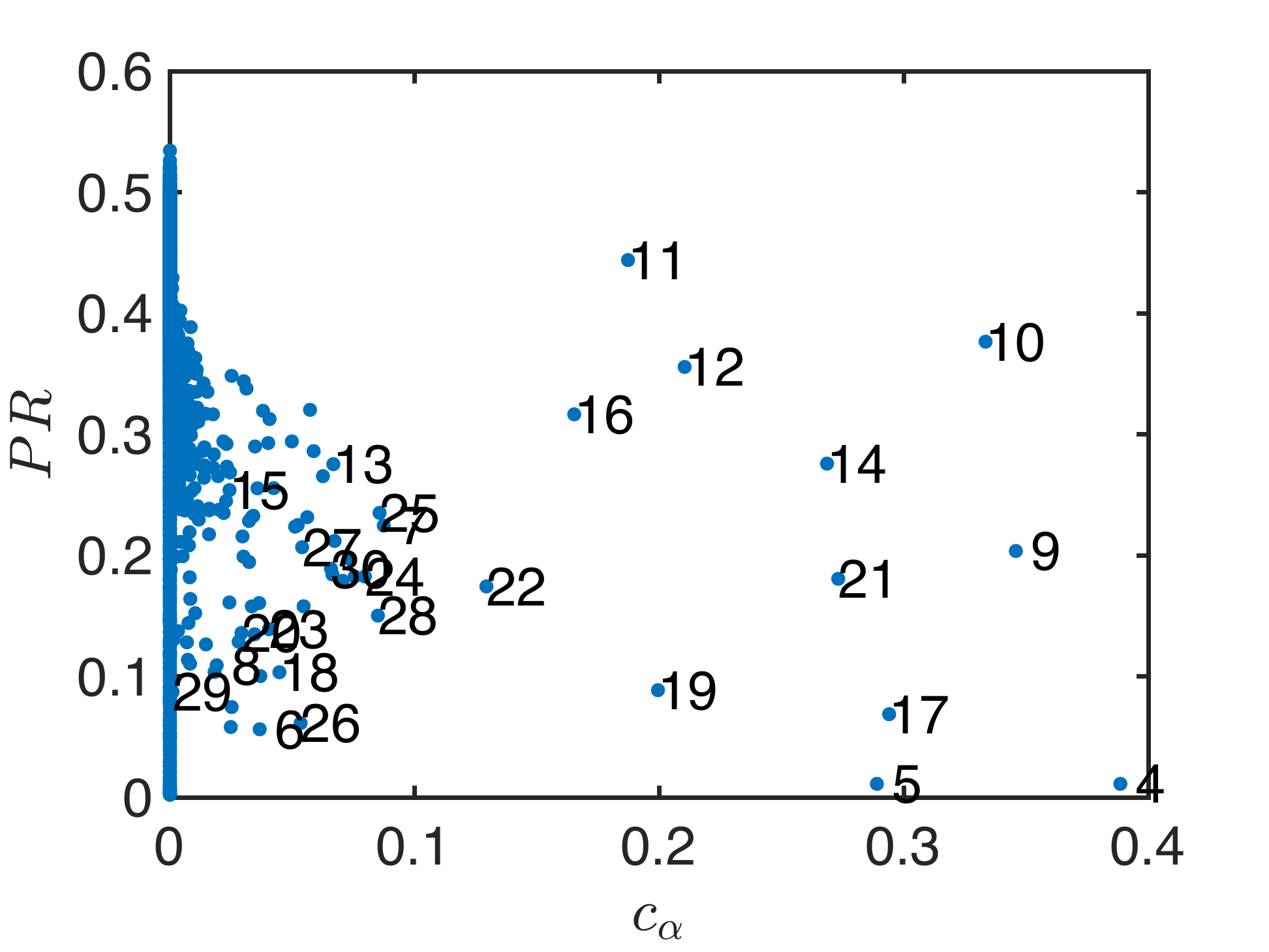}&
\includegraphics[scale=0.3]{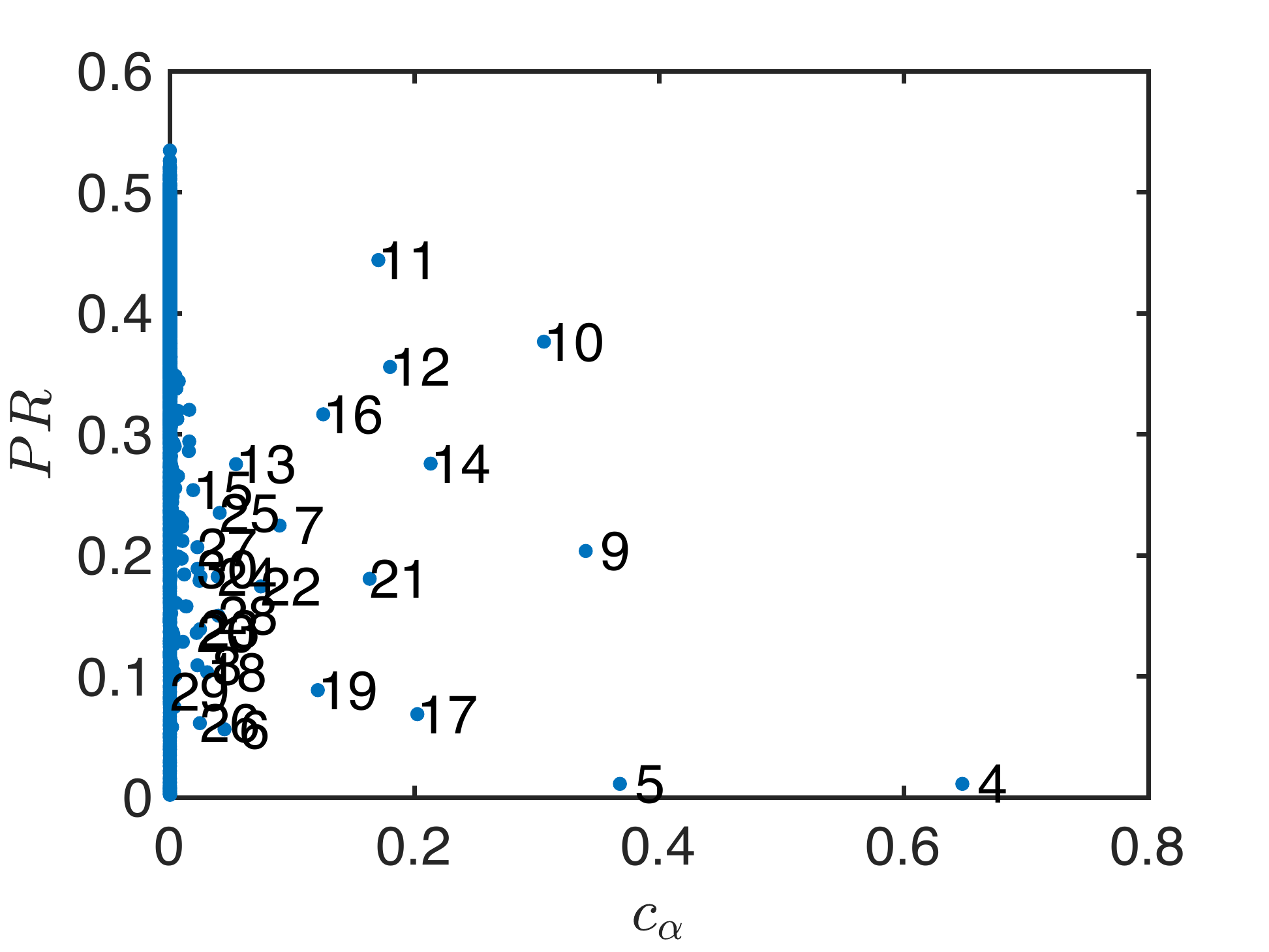}&
\includegraphics[scale=0.3]{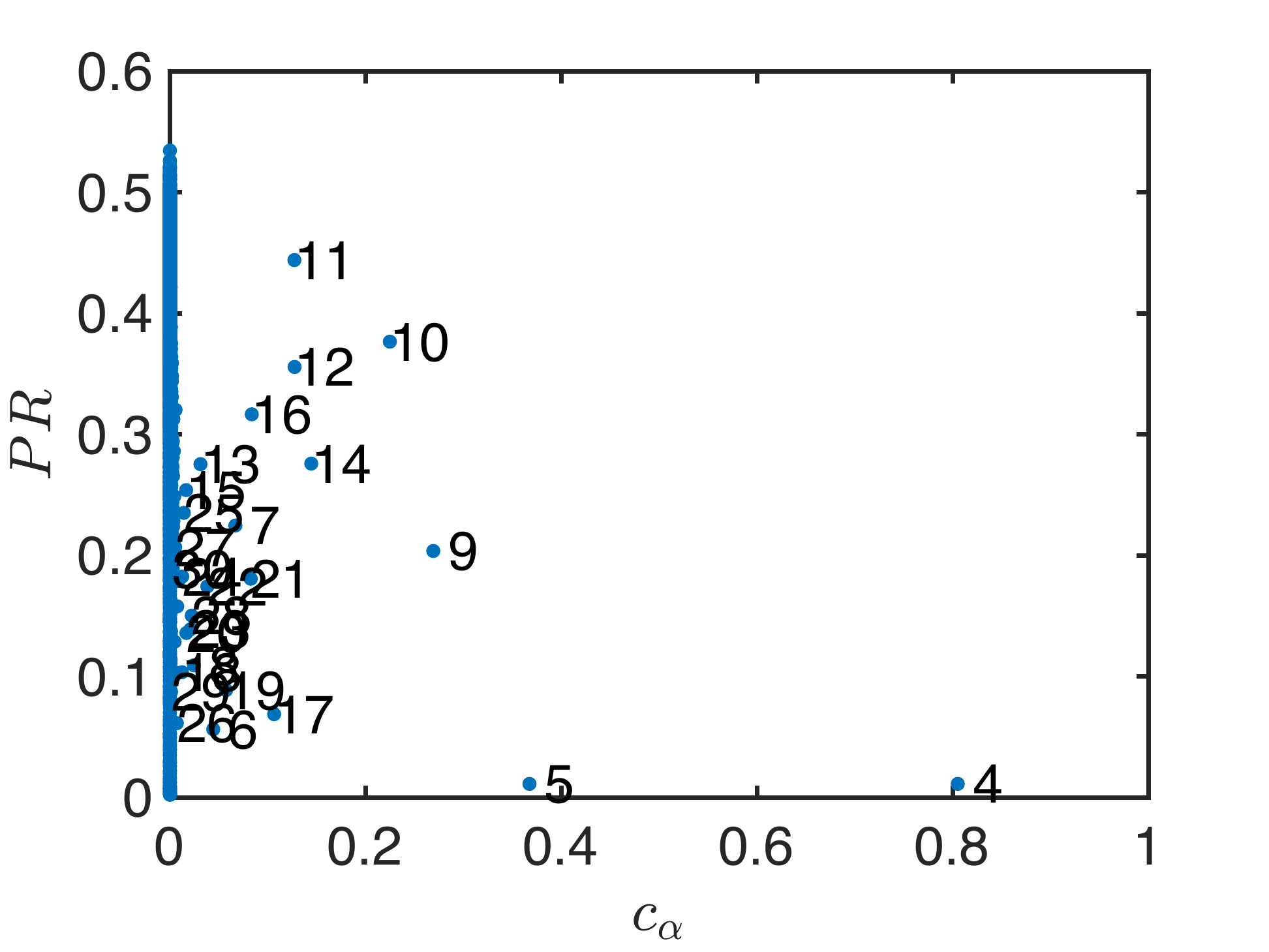}
\end{tabular}
\caption{Normalized scalar product $c_\alpha$ of the linear vibrational modes with $\ur_{12}$ versus their participation ratio, for different number $N_b$ of images in the NEB, for a selected active TLS (energy splitting $E=8.54\times10^{-4}$) for $T_f=0.062$. 
The modes are numbered starting from 4, mode 4 being thus the softest mode.
}
\label{fig:Nbeads}
\end{figure*}

 Non-linear modes are defined as follows.
 From the theoretical point of view, in order to define a DW, one needs to use a polynomial of at least fourth-order along a given coordinate. 
 However, it has been observed that the expansion of the potential energy up to the third order around an energy minimum already yields a reasonable approach to disentangle soft quasi-localized modes from phonons~\cite{Edan2016}.
 In our analysis we thus consider cubic nonlinear normal modes, defined as vectors that minimize the energy barrier 
 in the third-order approximation~\cite{Edan2016}, i.e. vectors $\upi$ that satisfy 
 \begin{equation}\label{NLM}
M\upi=\frac{M:\upi \, \upi}{U:\upi \, \upi \,\upi}U:\upi \, \upi \ ,
\end{equation}
where a colon denotes the contraction of a Greek index.
In contrast to linear modes, it is difficult to find all the solutions of the non-linear Eq.~\eqref{NLM}. In our analysis, we thus find numerically only one particular solution,
by using the iterative scheme suggested in Ref.~\cite{Edan2016}. 
We start from an initial guess $\upi^0$ (to be specified below), and we iterate a recursive equation derived from Eq.~\eqref{NLM},
\begin{equation}\label{recurs}
\upi^{n+1}=\frac{M:\upi^n \upi^n}{U:\upi^n\upi^n\upi^n}M^{-1}U:\upi^n\upi^n \ ,
\end{equation}
to find $\upi^1$, then from $\upi^1$ to find $\upi^2$, and so on. After several iterations, the vector $\upi^n$ converges to a non-linear mode.
Of course such a procedure does not allow one to find the full set of non-linear modes of the system, and it is not guaranteed that the non-linear mode is the closest one to the initial guess. Nevertheless, with these caveats in mind, this is the algorithm that we will use in our analysis due to the difficulty of finding a complete set of non-linear modes. 
Examples of the real-space character of a linear and a non-linear mode are shown in Fig.~\ref{fig:configuration}.

\begin{figure*}[t]
\begin{tabular}{ccc}
$\ur_{12}$&$\ur_{AS}$&$\ur_{AB}$\\
\includegraphics[scale=0.21]{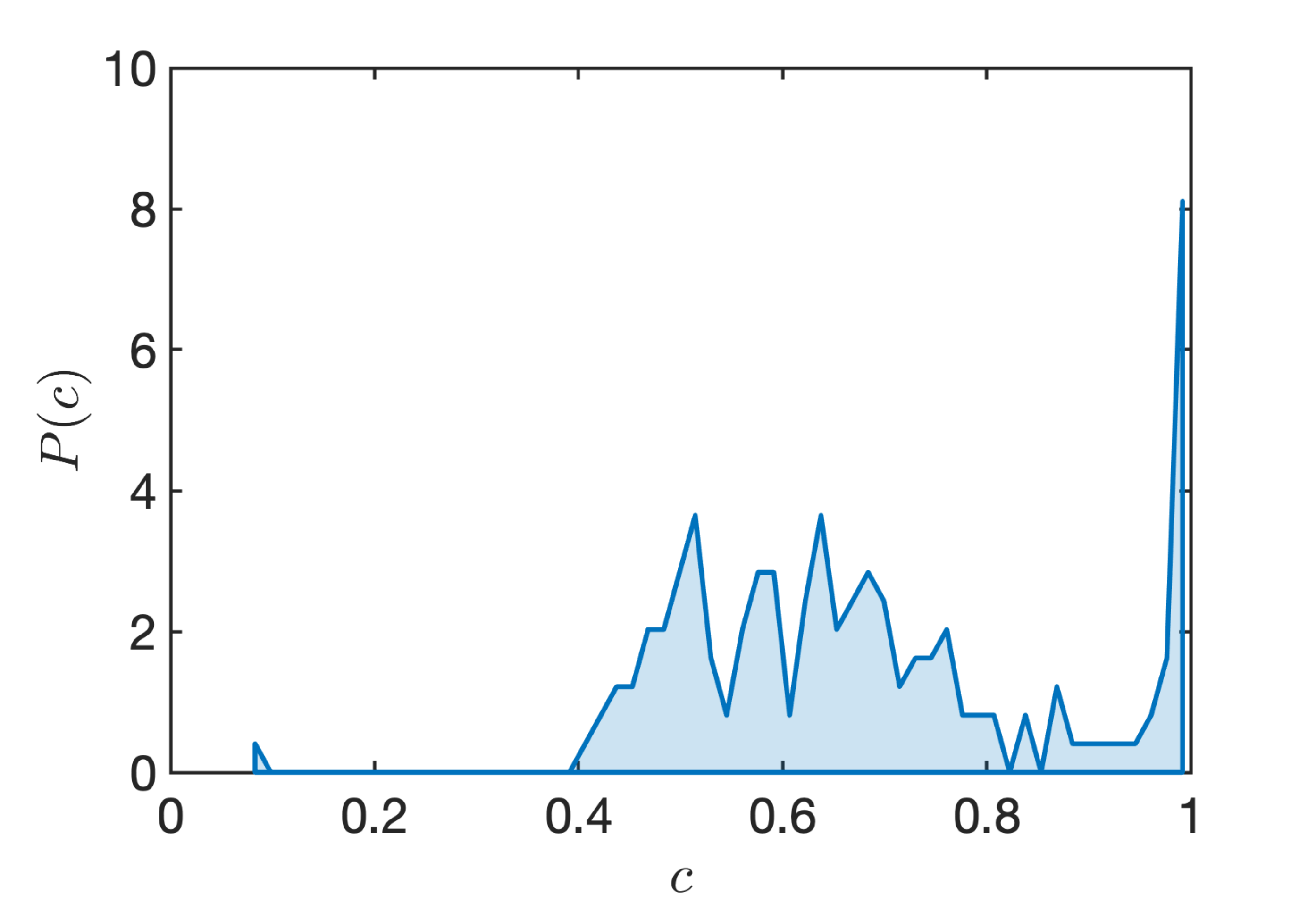}&
\includegraphics[scale=0.21]{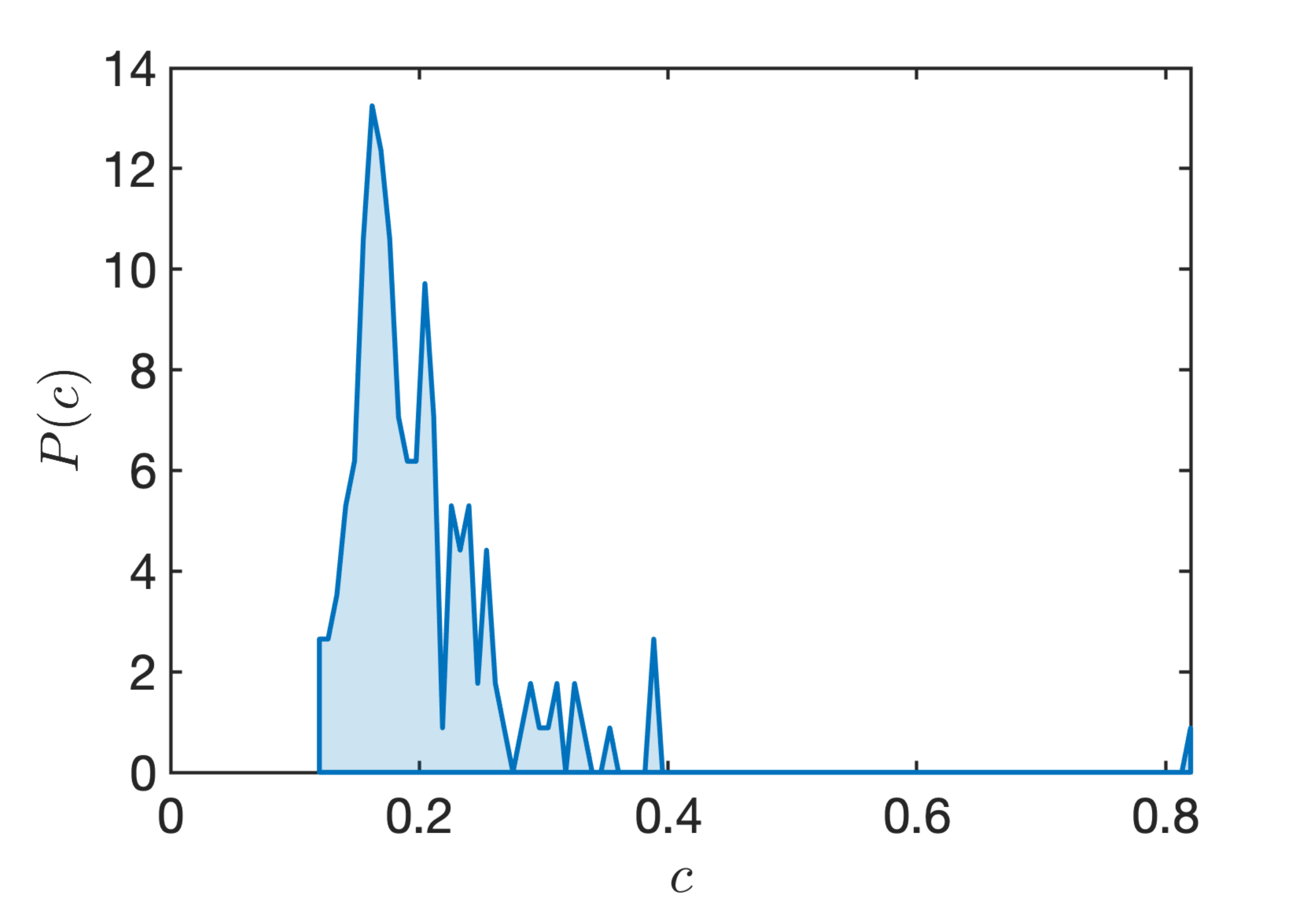}&
\includegraphics[scale=0.21]{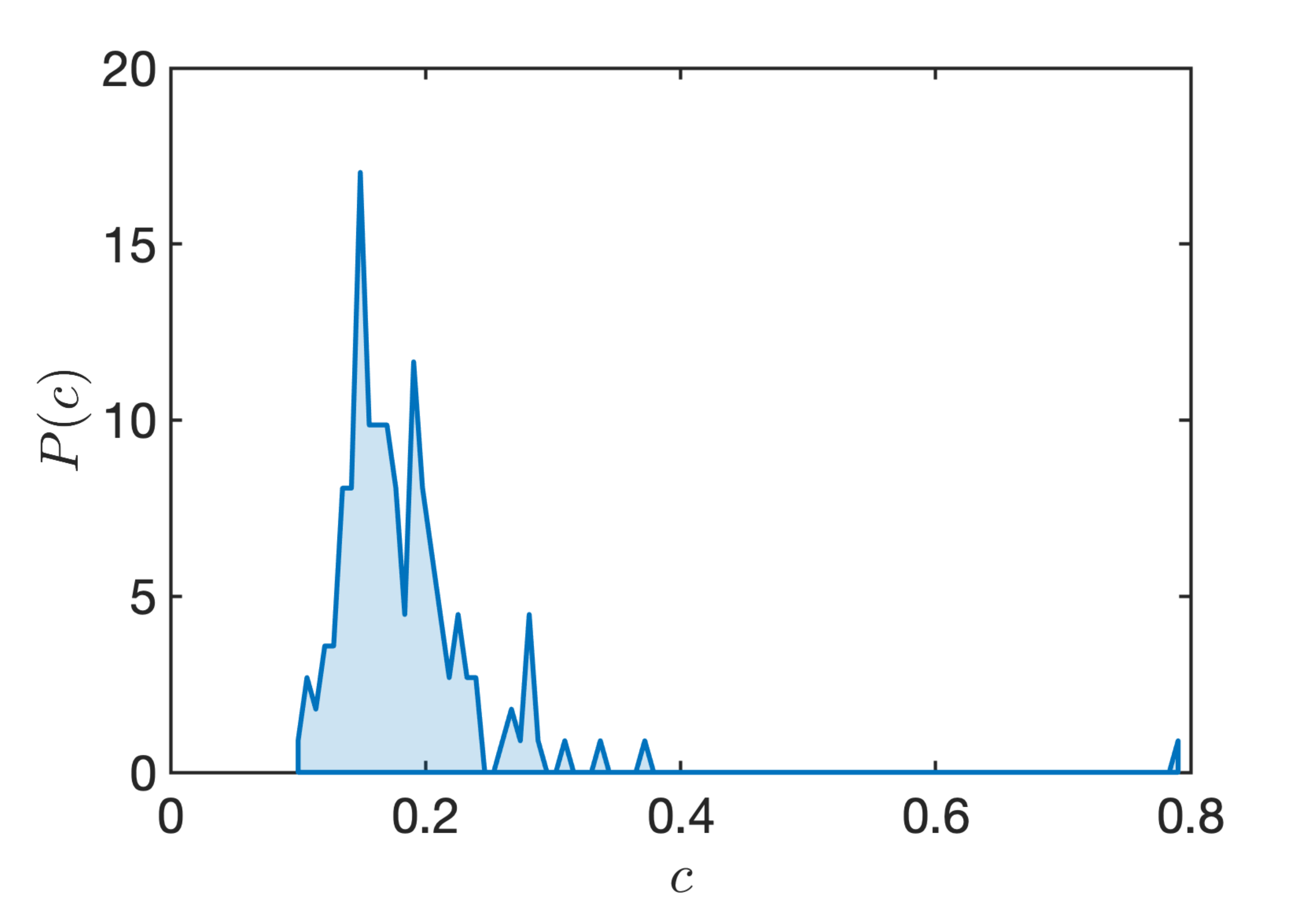}
\end{tabular}
\caption{Distribution over DW potentials of the largest mode projection, $c=\max_\alpha c_\alpha$, calculated over all modes for $\ur_{12}$, $\ur_{AS}$ and $\ur_{AB}$.
Here we used $N_b=600$, and since the NEB calculation is expensive, we have used a smaller subset of all DWs, only selecting TLS with energy splitting $E<10^{-3}$ and participation ratio of the DW transition $PR_{DW}<5$ (note that the $PR$ for DWs is normalized differently than for normal modes, see \cite{Khomenko2020}) for $T_f=0.062$.
}
\label{fig:r12modes} 
\end{figure*}

\section{Results}

We now report the results of the calculations described above, for all DWs in the data set obtained in Ref.~\cite{Khomenko2020}. 
 For $T_f=0.062, \, 0.07, \, 0.092$, there are 14202, 21109 and 117339 DWs, respectively. 
Of these, $61, 291, 1008$ are active TLS using Argon units, and $28, 46, 248$ are active using NiP units, respectively.

As a measure of overlap between a displacement field $\ur$ and a normal mode $\uv$, we will focus on two quantities: 
the simple normalized scalar product of $3N$-dimensional vectors, 
\begin{equation}\label{eq:cdef}
c(\uv)=\frac{ |\ur\cdot \uv| }{ |\ur||\uv|  } \ ,
\end{equation}
and the scalar product of $N$-dimensional vectors obtained by collecting the absolute values of particle displacements,
\begin{equation}\label{eq:adef}
a(\uv)=\frac{\sum_{i=1}^N |\mathbf{r}_i| |\mathbf{v}_i|}
{
\sqrt{
\sum_{i=1}^N |\mathbf{r}_i|^2
\sum_{i=1}^N |\mathbf{v}_i|^2
}
} \ ,
\end{equation}
where $\mathbf{r}_i$ and $\mathbf{v}_i$ are the displacements of particle $i$ in the $\ur$ and $\uv$ vectors. 
The parameter $a$ ignores the polarization of vectors and compares only the mobility of particles, and hence is an 
analog of the ``softness'' field used in \cite{wang2019low} for a single mode.

\subsection{Tangent vectors are parallel to a soft QLM}
\label{sec:r12}

For a given energy minimum, we can compute all the linear modes. Their participation ratio as a function of frequency is given
in Fig.~\ref{fig:local}, and an example of a QLM is given in Fig.~\ref{fig:configuration}. 

Our first result is that,
 in the limit of an infinite number of NEB images, ($N_b\to\infty$), when the path becomes continuous, the tangent vector $\ur_{12}$ to the minimal energy path in minimum $A$
 coincides with one of the softest linear modes in the same minimum. 
To prove this, for a selected DW, 
we show in Fig.~\ref{fig:Nbeads} the scalar product $c_\alpha$ between $\ur_{12}$ and all the vibrational modes in minimum $A$, here
labeled by $\alpha = 1,\cdots, 3N$. Note that the first three modes are trivial zero modes related to translational invariance, hence the non-trivial modes are labeled
by $\alpha = 4,\cdots, 3N$.
We clearly see that, upon increasing $N_b$ from 40 to 600, the overlap with mode 4, which is a QLM, increases while all the other overlaps decrease. Hence, we conclude that in the
limit $N_b\to\infty$, $\ur_{12}$ becomes essentially parallel to a soft QLM. 
Note that most of our simulations have been conducted with $N_b=40$, and that increasing the number of beads to $N_b=600$ makes the NEB calculation very computationally expensive, forcing us to restrict this investigation to a small number of DWs.

To provide further support for this statement, in Fig.~\ref{fig:r12modes} we report the probability distribution of the maximum overlap, $c = \max_\alpha c_\alpha$, between linear modes
and $\ur_{12}$, over a subset of DW potentials, as described in the figure caption. We observe that in all cases, $c>0.4$, and that in most cases $c$ is quite close to one, which confirms that the tangent vector is indeed parallel to a linear mode. Note that the results of Fig.~\ref{fig:r12modes} are for $N_b=40$, and we expect $c$ to increase upon increasing $N_b$.

The fact that $\ur_{12}$ is parallel to a linear mode, typically a soft QLM, implies that in real space there is always a soft QLM whose localized core is close to the particles that move the most in the DW. Hence, soft QLM are good predictors of the spatial location of DW potentials. 
 
\subsection{The frequency in a minimum is anti-correlated with the tunnel splitting}
 
 \begin{figure}[t]
   \includegraphics[width=\columnwidth]{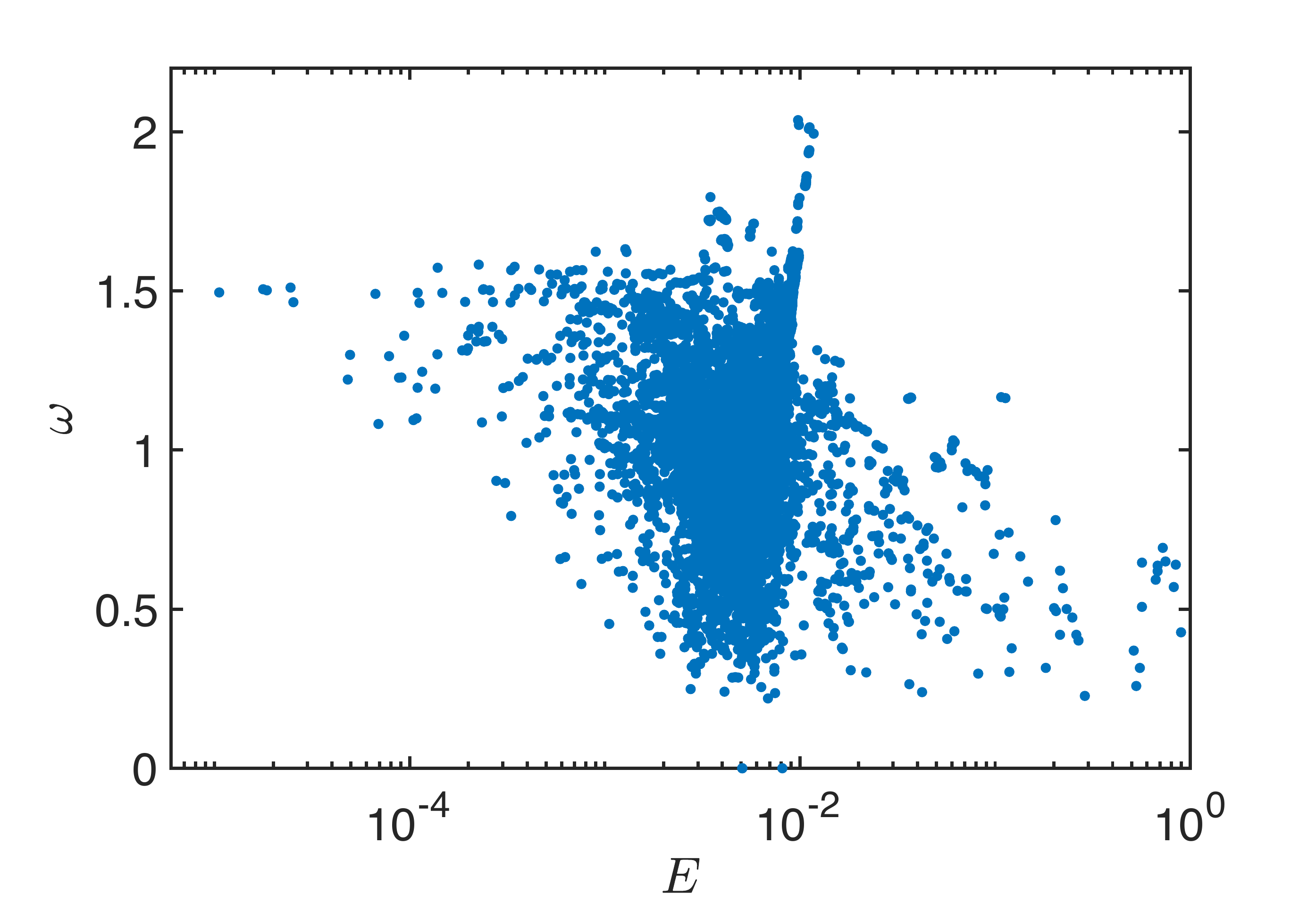}
\caption{Scatter plot of the tunnel splitting $E$ versus the frequency $\omega$ in minimum $A$, for all DWs found at $T_f=0.062$}
\label{fig:omegavsE}
\end{figure}

 Another interesting observation concerns the relation between the curvature $\omega$ of the energy profile along the transition path 
 (which, as discussed in section~\ref{sec:r12}, coincides with the frequency of a soft QLM) and the tunnel splitting $E$ associated to the DW.
 In Fig.~\ref{fig:omegavsE} we report a scatter plot of these two quantities, which shows a marked anticorrelation. 
 
We thus conclude that, although the vector $\ur_{12}$ is strongly associated with a soft linear mode, 
its frequency is not among the softest, and in particular the DWs with lowest splitting are associated to relatively higher frequencies. This behavior likely arises due to the fact that active TLS typically display a symmetric DW profile, with a relatively high barrier and hence a relatively high frequency of the two wells.

 \subsection{Linear modes are poor predictors of the minimal energy path curvature}

\begin{figure*}[t]
   \begin{tabular}{ccc}     
      \includegraphics[width=0.33\textwidth]{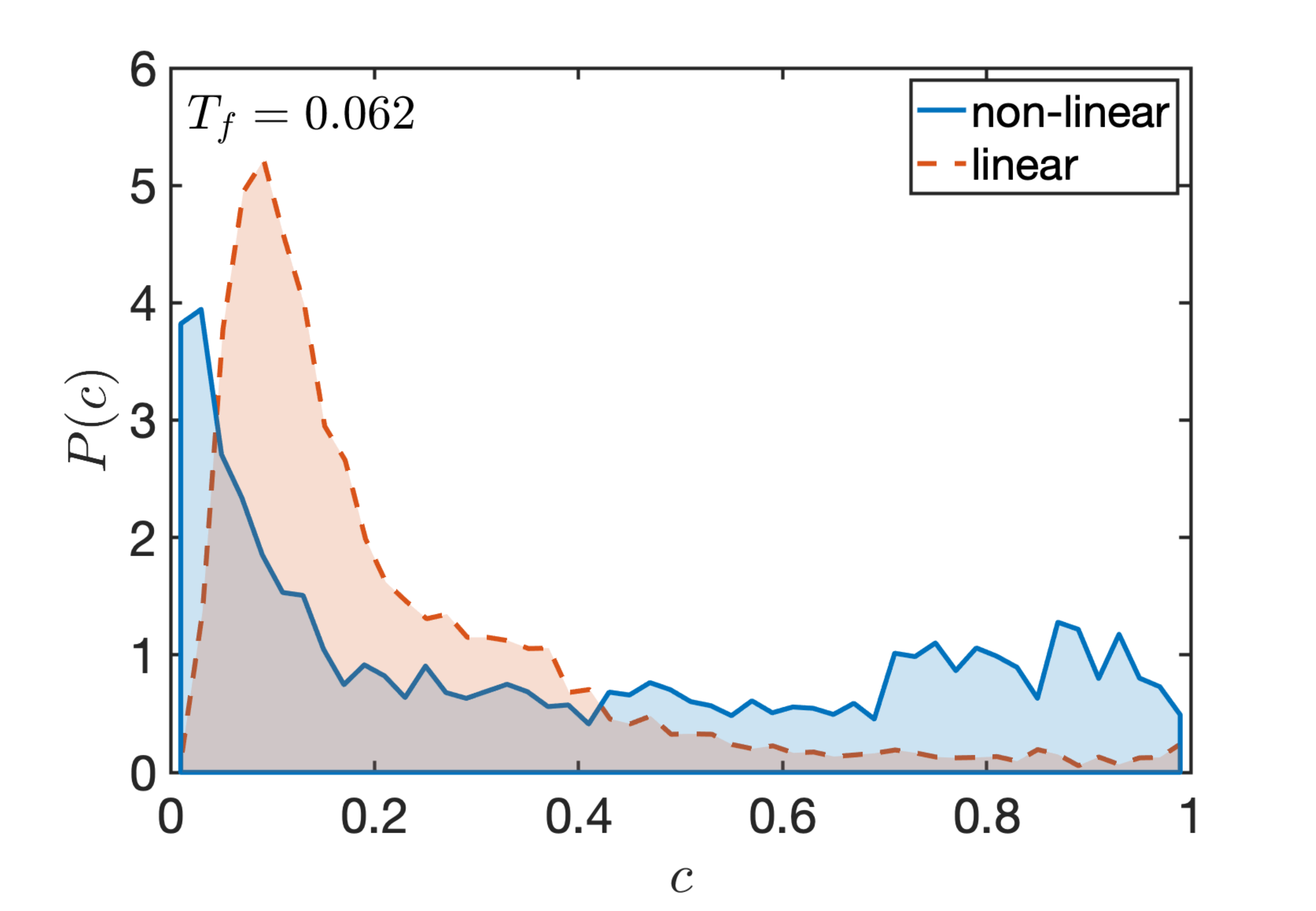}&
      \includegraphics[width=0.33\textwidth]{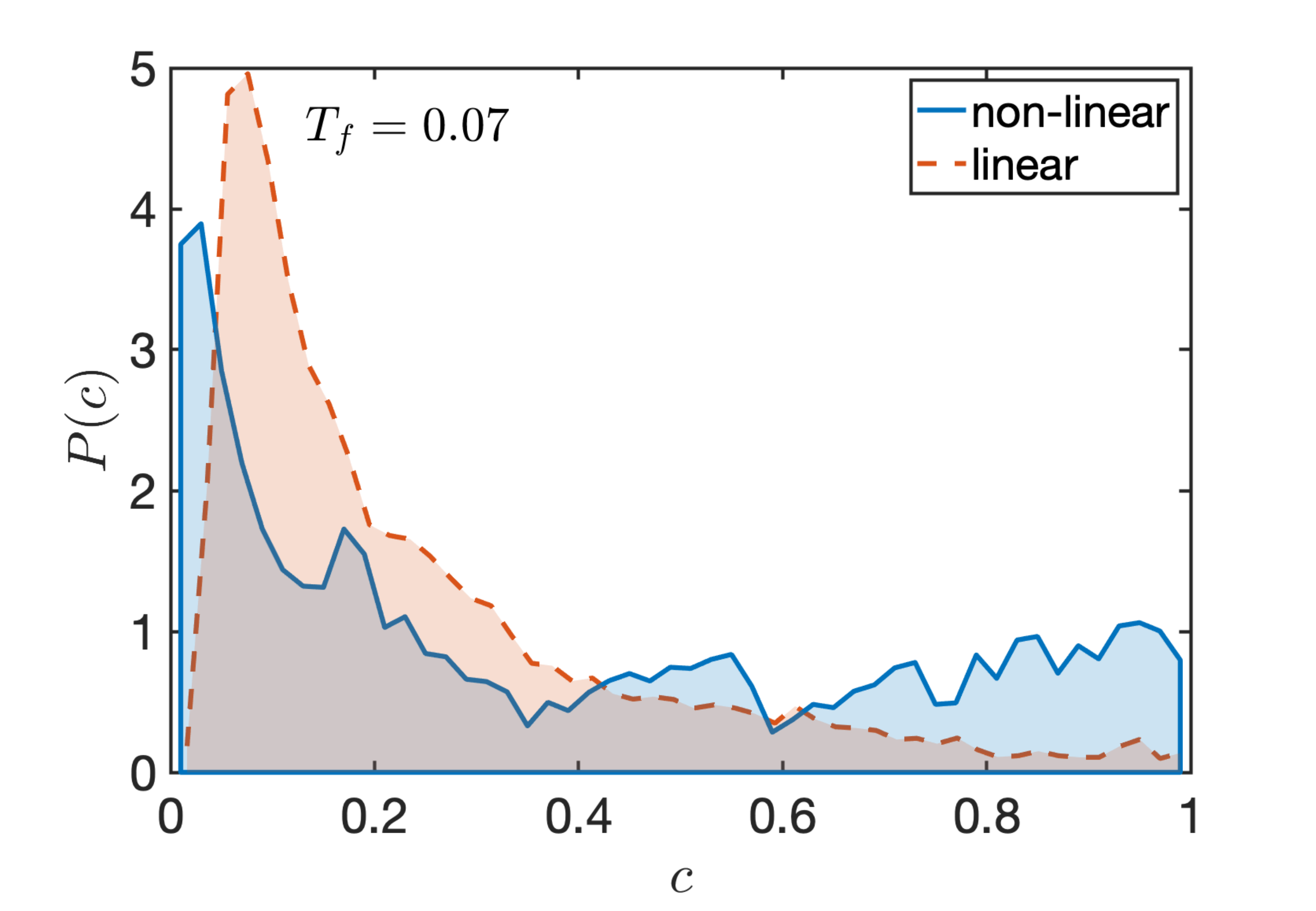}&
\includegraphics[width=0.33\textwidth]{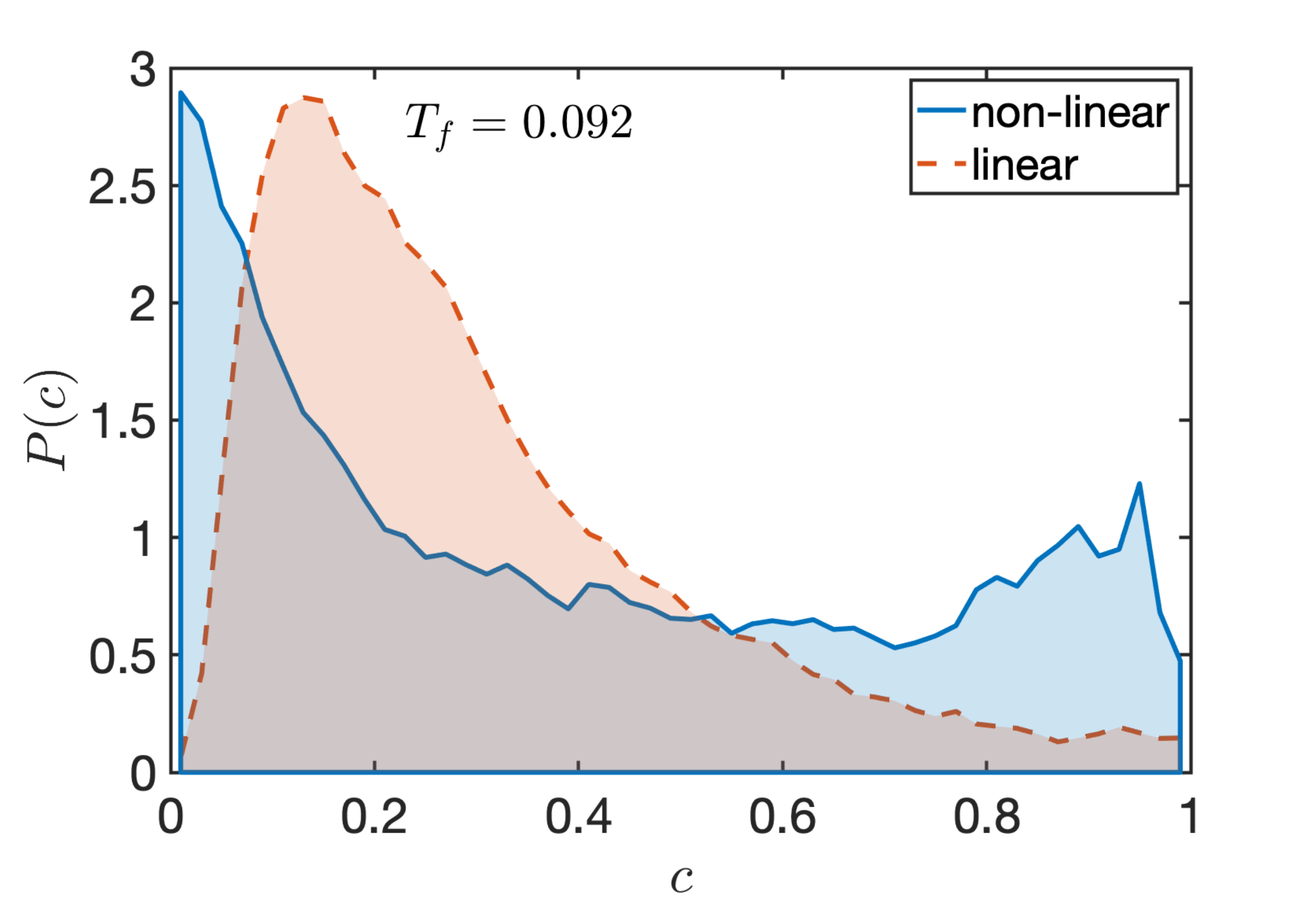}\\
      \includegraphics[width=0.33\textwidth]{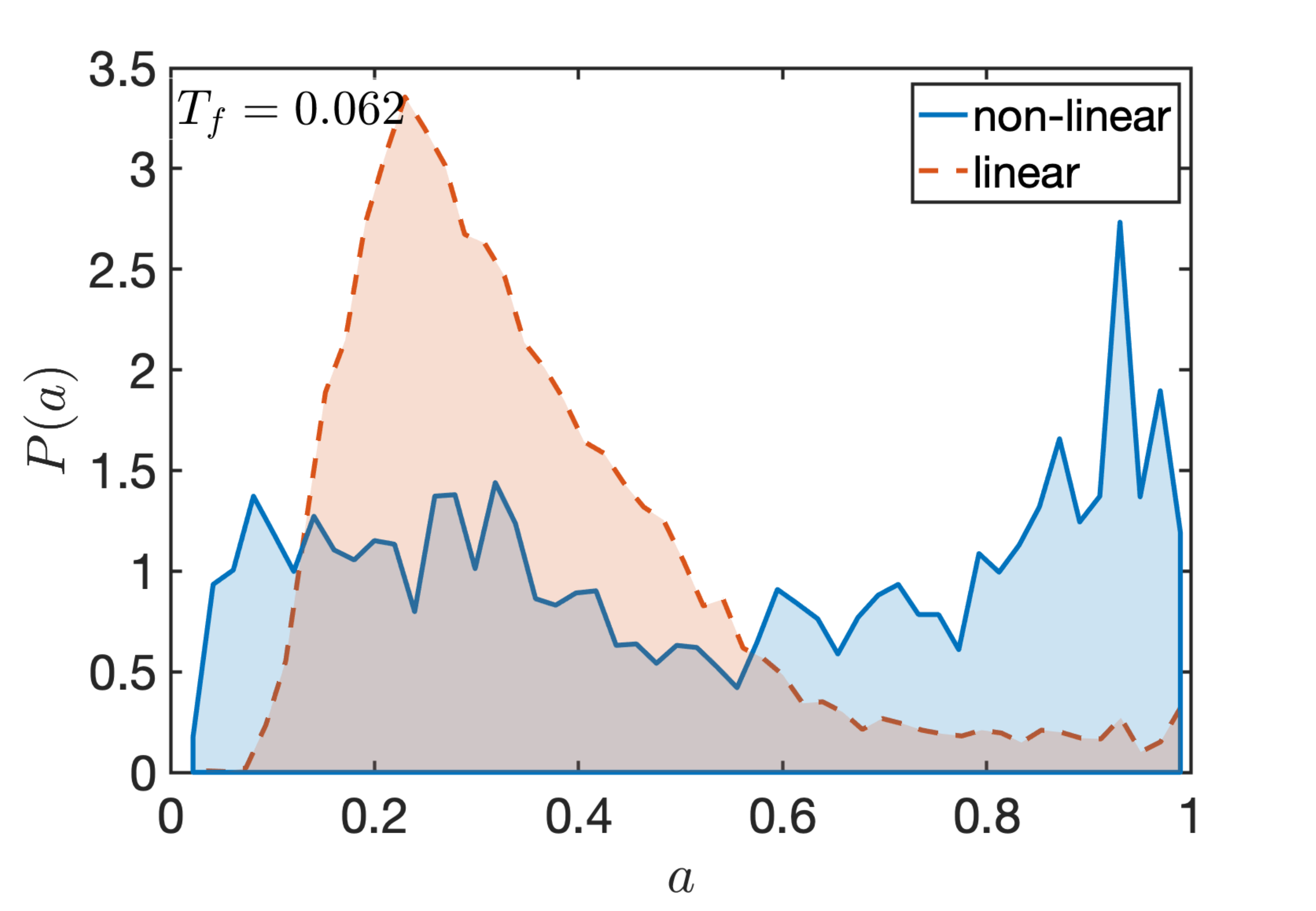}&
      \includegraphics[width=0.33\textwidth]{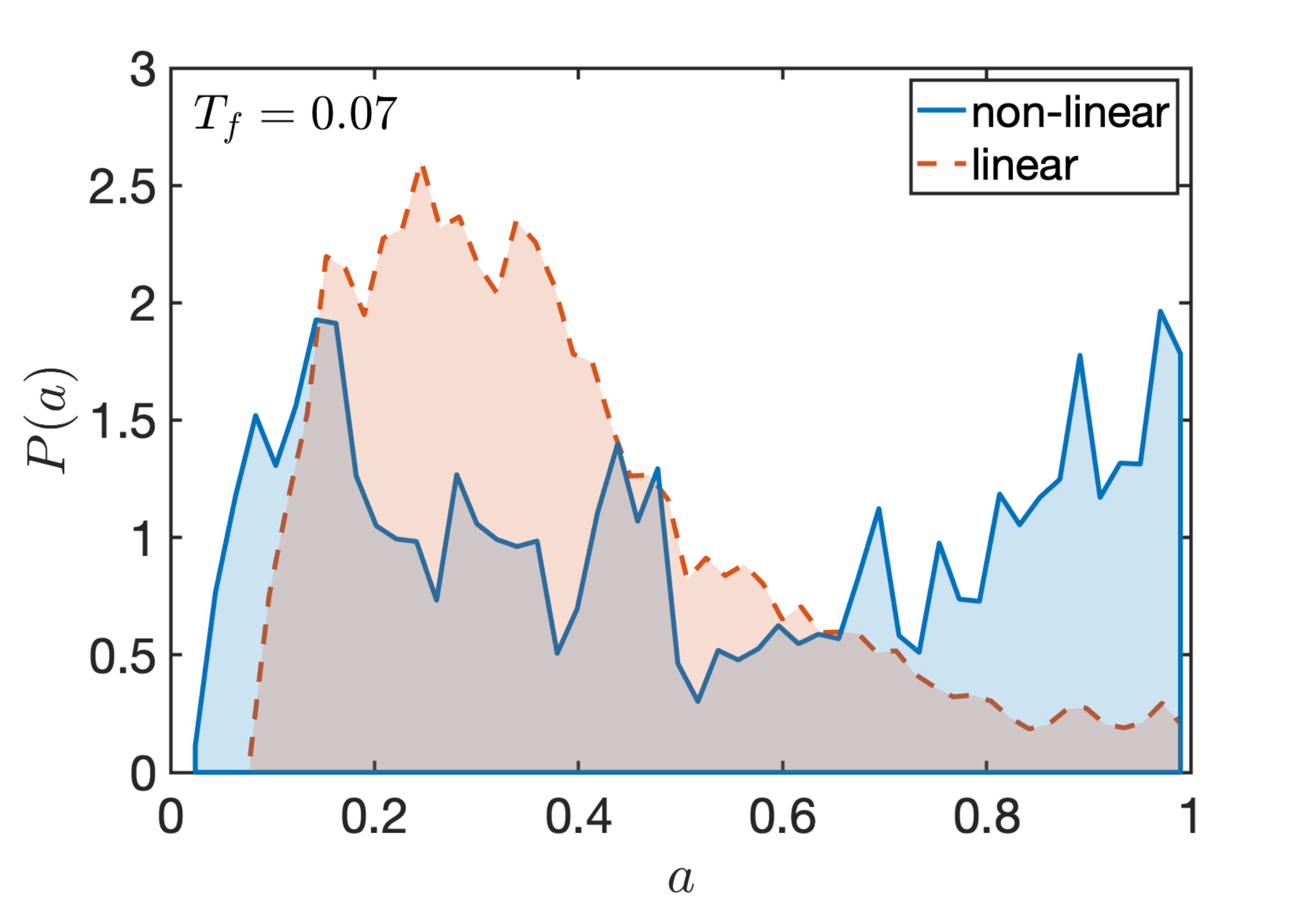}&
      \includegraphics[width=0.33\textwidth]{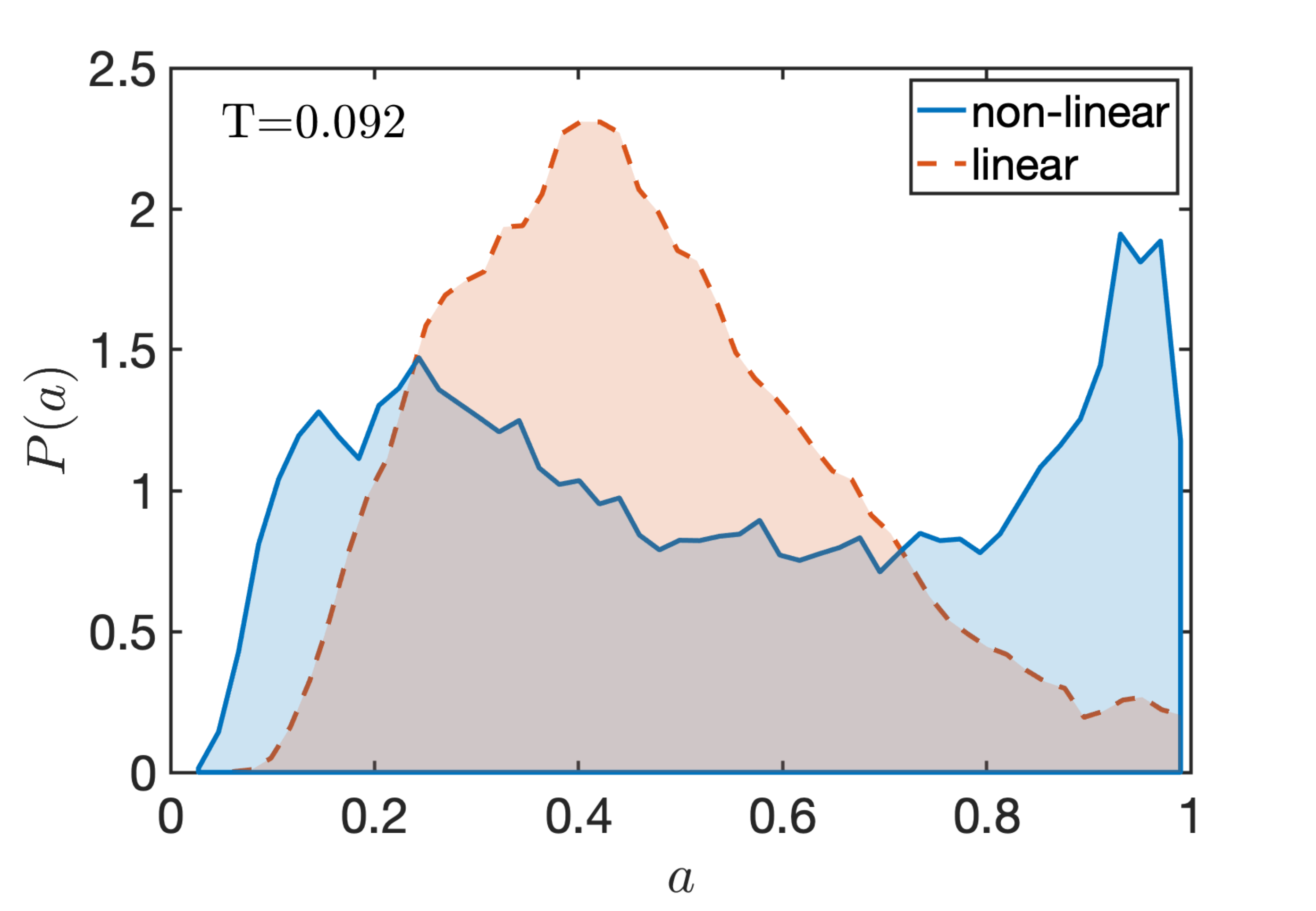}
    \end{tabular}
  \caption{Probability distribution of overlap coefficients $c$ and $a$ for linear and non-linear modes, obtained from a recursive procedure, starting from $\ur_{AB}$ as initial guess. Data are for the three preparation temperatures $T_f$ and for the full set of DWs.}
\label{fig:statistics}   
\end{figure*} 

\begin{figure}[t]
      \includegraphics[width=\columnwidth]{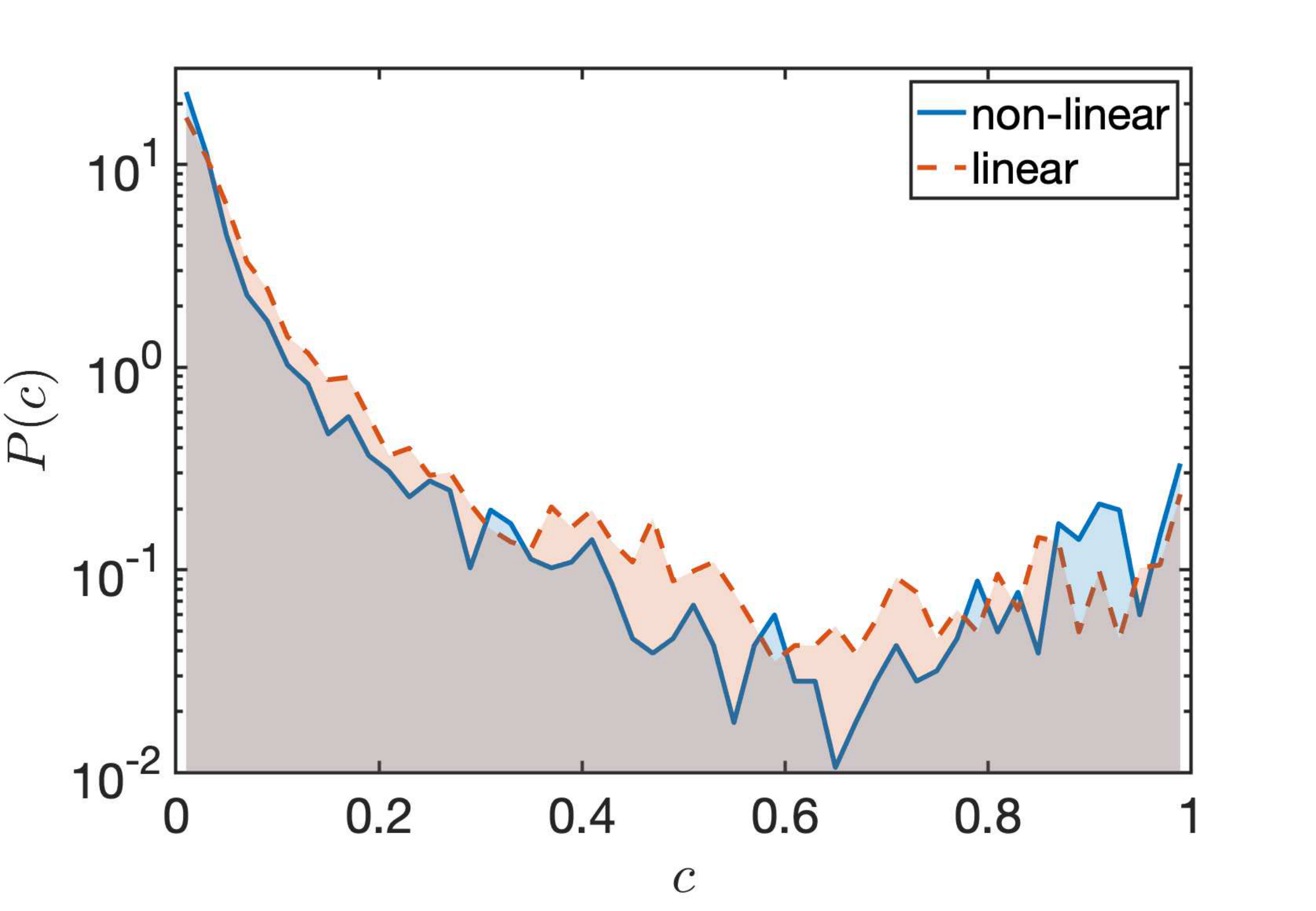}
  \caption{Overlap of the softest localized linear modes, and of the non-linear modes that are constructed using those linear modes as an initial condition, with the displacement field $\ur_{AB}$. Data are for $T_f=0.062$ and for the full set of DWs.}
\label{fig:rand}   
\end{figure}

We now discuss whether the linear modes of minimum $A$ are good predictors of the transition path associated with a DW, as encoded
by the minimum-to-saddle displacement $\ur_{AS}$ and by the minimum-to-minimum displacement $\ur_{AB}$.
In Fig.~\ref{fig:r12modes} we compare the statistics of the maximum overlap coefficient $c=\max_\alpha c_\alpha$ of linear modes with $\ur_{12}$, $\ur_{AS}$, and $\ur_{AB}$. 
From these figures we clearly see that the values of $c$ for $\ur_{AS}$ and $\ur_{AB}$ are much smaller than for $\ur_{12}$.
We thus conclude that the minimal energy path between two energy minima, which (as shown in section~\ref{sec:r12}) tends to start along one of the softest modes locally around each minimum, markedly changes its direction upon approaching the saddle point. We find that linear modes are poor predictors of this change of direction.

\subsection{Non-linear modes are better correlated with the minimum energy path curvature}

We next consider whether non-linear modes can be better predictors of the direction of the minimum energy path around the saddle point $S$ or the arrival minimum $B$.
Because most of the data we will show are qualitatively similar for $\ur_{AB}$ and $\ur_{AS}$, we will focus on the former for the rest of this section.

In order to find the closest non-linear mode to the minimum energy path, we use $\upi_0 = \ur_{AB}/|\ur_{AB}|$ as an initial guess for
the iterative procedure in Eq.~\eqref{recurs}, and we iterate until convergence to the corresponding non-linear mode.
We then compute the overlap coefficients $c$ and $a$ between the non-linear mode and $\ur_{AB}$, defined respectively in Eq.~\eqref{eq:cdef} and Eq.~\eqref{eq:adef}.
In order to provide a direct comparison with linear modes, we find
the linear mode that has the maximum overlap ($a$ or $c$) with $\ur_{AB}$.
The statistics of $a$ and $c$ is shown in Fig.~\ref{fig:statistics} for the full set of available
DWs at the three $T_f$ values. 
From these plots, one can see that non-linear modes generally have a much stronger overlap with $\ur_{AB}$ than do the linear modes, but nevertheless, for a large fraction of DWs, the overlap remains small even for non-linear modes. 
It is very important to stress that in this analysis we used the {\it a priori} known information encoded in $\ur_{AB}$ as an initial guess to search for the closest non-linear mode to the reaction path. Hence, we expect the values reported in Fig.~\ref{fig:statistics} to provide an upper bound on the possible overlaps. 

We have repeated the same analysis, without assuming any {\it a priori} knowledge about the second minimum. We first diagonalize the Hessian matrix in minimum $A$, and identify the softest mode, which we use as an initial guess for the non-linear mode search. We then compare the resulting non-linear mode with $\ur_{AB}$ by computing the overlap $c$, shown in 
Fig.~\ref{fig:rand} for the linear mode used as initial guess and for the corresponding non-linear mode. We observe in this case that linear and non-linear modes have comparably poor predictive power. Of course, we cannot exclude that there is another DW starting from minimum $A$ and connecting to another minimum $B'$, which might be better correlated with these modes, although this scenario seems unlikely.

Finally, we investigated whether the energy profile along the non-linear modes we found using this procedure, i.e.
$v(s)=V(\ur_{A} + s \upi)$, displays a DW shape, and we did not find any DW in approximately $95\%$ of cases. This is consistent with results reported in~\footnote{
Edan Lerner, private communication. See the Lectures at the workshop {\it Recent progress in glassy systems}, Les Houches, February 2020 for related results. Slides available at \url{https://sites.google.com/view/leshouches2020}.
}, and illustrates the complexity of the energy landscape in which TLS reside.

\section{Conclusions} 

In this paper, we have investigated the relationship between QLM and TLS {\it in silico} in a model glass, exploiting the TLS library constructed in~\cite{Khomenko2020}. We find that soft QLM are generally associated with the initial direction of the minimum energy path connecting two minima, and as a consequence, DW potentials are spatially located close to a soft QLM. However, the frequency of the QLM is anticorrelated with the tunnel splitting associated to the DW, hence TLS are typically not associated to the softest modes, which on the contrary are expected to be responsible for plasticity~\cite{KLLP10,david_huge_collaboration}. We conclude that QLM with properly tuned frequency could serve as good predictors of the spatial location of TLS.  However, we also find that the minimal energy path is strongly curved within a high-dimensional space, in such a way that the saddle point and the secondary minimum are uncorrelated with the direction of the initial tangent vector. We find that linear modes are poor predictors of the minimum-to-saddle or minimum-to-minimum directions. 

We have also considered non-linear cubic modes~\cite{Edan2016} and find that one of these modes is often well correlated with the minimum-to-minimum direction $\ur_{AB}$. However, locating this individual mode is difficult: if the search is initialized with $\ur_{AB}$ itself, convergence to the correct mode is facile. If, on the contrary, the search is initialized in a soft linear mode, convergence to the correct non-linear mode does not occur. We conclude that in absence of some prior information about the direction of $\ur_{AB}$, it is difficult to predict the more global displacement field associated with TLS via either linear or non-linear modes. 

The problem of finding good structural predictors for TLS thus remains somewhat open. It is possible that better search strategies could exploit the information contained in linear or non-linear modes more efficiently. Machine learning techniques~\cite{Sussman10601,bapst2020unveiling} might be able to exploit this information (and perhaps additional structural information) to achieve better performance at contact prediction. Exploring this possibility is a clear direction for future work.

\acknowledgements

We would like to thank Ludovic Berthier, Eran Bouchbinder, Wencheng Ji, Edan Lerner, Felix-Cosmin Mocanu, Corrado Rainone, Camille Scalliet, Pierfrancesco Urbani, Matthieu Wyart for useful discussions and Ludovic Berthier and Camille Scalliet for assistance and collaboration at the early stages of this project.

This project has received funding from the European Research Council (ERC) under the European Union's Horizon 2020 research and innovation programme (grant agreement n. 723955 - GlassUniversality), and it was supported by a grant from the Simons Foundation (\#454951 David Reichman, \#454955, Francesco Zamponi).
This work was granted access to the HPC resources of MesoPSL financed
by the Region Ile de France and the project Equip@Meso (reference
ANR-10-EQPX-29-01) of the programme Investissements d'Avenir supervised
by the Agence Nationale pour la Recherche. 

\bibliography{main.bib}

\clearpage

\appendix
\renewcommand{\thefigure}{S\arabic{figure}}
\setcounter{figure}{0}
\begin{widetext}

\end{widetext}

\end{document}